\documentclass[twocolumn,showpacs,aps,superscriptaddress,prl]{revtex4}
\usepackage{amssymb,amsmath}
\usepackage[dvips]{graphicx}

\usepackage[dvipdfm,letterpaper]{hyperref}

\begin{document}

\title{An Anomalously Elastic, Intermediate Phase  in Randomly
Layered Superfluids, Superconductors, and Planar Magnets}

\author{Priyanka Mohan}
\affiliation{Department of Physics, Indian Institute of Technology
Madras, Chennai 600036, India}
\author{Paul M.\ Goldbart}
\affiliation{Department of Physics, University of Illinois at
Urbana-Champaign, 1110 West Green St., Urbana, IL 61801, USA}
\author{Rajesh Narayanan}
\affiliation{Department of Physics, Indian Institute of Technology
Madras, Chennai 600036, India}
\author{John Toner}
\affiliation{Department of Physics and Institute of Theoretical
Science, University of Oregon, Eugene, OR 97403, USA}
\author{Thomas Vojta}
\affiliation{Department of Physics, Missouri University of Science
and Technology, Rolla, MO 65409, USA}

\begin{abstract}
We show that layered quenched randomness in planar magnets leads to an unusual
intermediate phase between the conventional ferromagnetic low-temperature and
paramagnetic high-temperature phases.  In this intermediate phase, which is part of the
Griffiths region, the spin-wave stiffness perpendicular to the random layers displays
anomalous scaling behavior, with a continuously variable anomalous exponent, while the
magnetization and the stiffness parallel to the layers both remain finite.  Analogous
results hold for superfluids and superconductors. We study the two phase transitions into
the anomalous elastic phase, and we discuss the universality of these results, and
implications of finite sample size as well as possible experiments.
\end{abstract}


\maketitle


The macroscopic behavior of many-particle systems is often sensitive to quenched
disorder. For example, at zero-temperature quantum phase transitions, the interplay of quantum and
disorder fluctuations gives rise to exotic phenomena, such as quantum Griffiths
singularities \cite{ThillHuse95,GuoBhattHuse96,RiegerYoung96}, infinite-randomness
critical points \cite{Fisher92,Fisher95}, and smeared transitions
\cite{Vojta03a,Vojta06}. The main reason for these strong effects of disorder is the presence of perfect disorder correlations in the
\emph{imaginary-time} dimension, which becomes infinitely extended at zero temperature.
Thus, one is effectively dealing with infinitely  large impurities.

This suggests that strong disorder effects should also occur at classical
(thermal) phase transitions, if the disorder is perfectly correlated in one or more
\emph{space dimensions}. For example, the McCoy-Wu model \cite{McCoyWu68,McCoyWu68a}, a
two-dimensional (2D) Ising model in which the disorder is perfectly correlated in one
dimension, shows an exotic transition, characterized by a smooth specific
heat but an infinite susceptibility over a range of temperatures. By using a
strong-disorder renormalization group, Fisher \cite{Fisher92,Fisher95} showed that
the critical point is of the infinite-randomness kind, and is accompanied by power-law
Griffiths singularities. Similar behavior was found in Heisenberg magnets
having 2D disorder correlations \cite{MohanNarayananVojta10}.

\begin{figure}
\includegraphics[angle=0,width=7.8cm,clip]{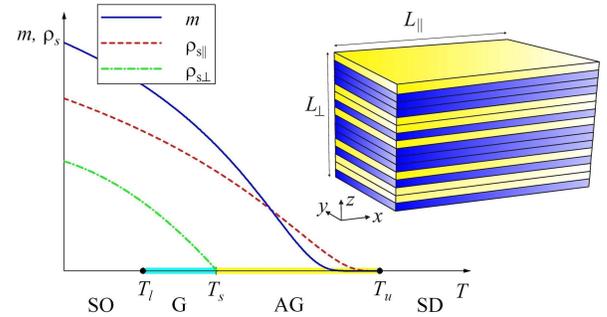}
\caption{(Color online) Schematic behavior of the magnetization $m$ and the stiffnesses
$\rho_{s,\parallel}$ and $\rho_{s,\perp}$ vs. temperature $T$ for a bounded disorder
distribution. SD and SO denote the conventional strongly disordered and ordered phases,
respectively. The Griffiths region (bounded by $T_u$ and $T_l$) consists of the
``non-anomalous'' (G) and the anomalous (AG) Griffiths phases. For an unbounded
distribution, $T_u\rightarrow\infty$. Inset: Randomly layered magnet or superconductor:
layers of two distinct materials are deposited in a random sequence.}
\label{Fig:layeredmagnet}
\end{figure}

In this Letter, we study thermal phase transitions exhibited by
randomly layered
3D superfluids, superconductors, and planar magnets, as sketched in Fig.\
\ref{Fig:layeredmagnet}.
All these systems are characterized by two-component order parameters of $U(1)$ or,
equivalently, $O(2)$ symmetry (representing the condensate wave function, Cooper pair
amplitude, and magnetization, respectively).

Couched in terms of the planar ferromagnet, our results can be summarized as follows:
The interplay of the layered randomness and the Kosterlitz-Thouless (KT)
\cite{KosterlitzThouless73} transitions in strongly
coupled multilayers (slabs) leads to an anomalously elastic intermediate phase as part of
the Griffiths region associated with the phase transition. In this anomalous Griffiths
phase, the magnetization $m$ and the spin-wave stiffness $\rho_{s,\parallel}$ parallel to
the layers are both nonzero (as in a conventional ferromagnet). However, the
stiffness $\rho_{s,\perp}$ \emph{perpendicular} to the layers vanishes, and the elastic
free energy exhibits anomalous scaling behavior. Specifically, free energy $\Delta F$
due to twisted boundary conditions (BCs) in the $z$ direction, forcing the spins on the
top face in Fig.\ \ref{Fig:layeredmagnet} to make an angle of $\Theta$ with those
on the bottom face, varies as
\begin{equation}
{\Delta F} \sim \Theta^2 L_\perp^{-z}
\label{eq:anom}
\end{equation}
with system size $L_\perp$. Here, $z(T)$ is a temperature dependent dynamical exponent
that varies continuously from $z=\infty$ at the upper Griffiths temperature $T_u$ (i.e.,
the boundary between the Griffiths region and the conventional paramagnet) to $z=1$ at
the temperature $T_s$ where a nonzero perpendicular stiffness first appears.

While anomalous elasticity of this type occurs in some disordered systems possessing {\it
uncorrelated} disorder (e.g., liquid crystals in aerogel \cite{RadzihovskyToner99}), in
those cases it is characterized by \emph{universal} values of the anomalous exponent $z$:
the non-universality and strong temperature dependence of $z$ that we find here are, to
the best of our knowledge, unique to systems having correlated disorder.

We also find unusual behavior at the two transition temperatures $T_u$
and $T_s$.  The magnetization $m$ is nonzero for all $T< T_u$ and shows
a double-exponential tail towards the nonmagnetic phase. Close to $T_u$, it takes the asymptotic form
\begin{equation}
\ln(m) \sim -\exp[a (T_u-T)^{-\nu}]~, \qquad (T\to T_u-) \label{eq:m-T}
\end{equation}
where $\nu \approx 0.6717$ \cite{CHPV06} is the correlation-length
critical exponent of a clean
3D planar (XY) magnet and $a$ is a nonuniversal constant.
If an external magnetic field $H$ is applied at temperatures $T \alt T_u$,
the magnetization vanishes with decreasing field more slowly than any
power,
\begin{equation}
\ln(m) \sim - \sqrt{|\ln(H)| (T_u-T)^{-\nu}}~,    \qquad (H\to 0)~. \label{eq:m-H}
\end{equation}
This relation applies for magnetizations larger than the
double-exponentially small value
given in (\ref{eq:m-T}).

The \emph{parallel} spin-wave stiffness (corresponding to a twist of the BCs in the $x$
or $y$ direction) $\rho_{s,\parallel}$ is nonzero for all $T<T_u$ and shows an
exponential tail of the form
\begin{equation}
\ln(\rho_{s,\parallel}) \sim (T_u-T)^{-\nu}~, \qquad (T\to T_u-)~.
\label{eq:rho_s_parallel}
\end{equation}
In contrast, the \emph{perpendicular} stiffness $\rho_{s,\perp}$ (corresponding to a twist of
the BCs in the $z$ direction) vanishes as $T_s$ is approached from below via
\begin{equation}
\rho_{s,\perp} \sim (T_s-T) ~, \qquad (T\to T_s^-)~. \label{eq:rho_s_perp}
\end{equation}


In the remainder of this Letter, we sketch the derivation of these results, compute
finite-size effects, and discuss possible experimental realizations. For definiteness, we
focus on the classical planar ferromagnet (i.e., the XY model) on a cubic lattice. The
Hamiltonian is given by
\begin{equation}
H = - \sum_{\mathbf{r}} J^{\parallel}_z \, (\mathbf{S}_{\mathbf{r}}
\cdot \mathbf{S}_{\mathbf{r}+\hat{\mathbf{x}}}
                                         +\mathbf{S}_{\mathbf{r}}
\cdot \mathbf{S}_{\mathbf{r}+\hat{\mathbf{y}}} )
     - \sum_{\mathbf{r}} J^{\perp}_z \, \mathbf{S}_{\mathbf{r}}
\cdot\mathbf{S}_{\mathbf{r}+\hat{\mathbf{z}}}
     .
\label{eq:Hamiltonian}
\end{equation}
Here, $\mathbf{S}_{\mathbf{r}}$ is a two-component unit vector at lattice site
$\mathbf{r}$, and  $\hat{\mathbf{x}}$, $\hat{\mathbf{y}}$, and
$\hat{\mathbf{z}}$ are
the unit vectors in the coordinate directions. The exchange interactions within
the layers, $J^{\parallel}_z$, and between the layers, $J^{\perp}_z$,
are both positive
and independent random functions of the perpendicular coordinate $z$.
For simplicity, we take all $J_z^\perp \equiv J^\perp$ and draw the $J_z^\parallel$
from a binary distribution ($J_u > J_l$)
\begin{equation}
P(J^{\parallel})=(1-c)\, \delta(J^{\parallel} - J_u) + c\,
\delta(J^{\parallel} - J_l)~,
\label{eq:binary_distrib}
\end{equation}
where $c$ is the concentration of ``weak'' layers.


Let us discuss the planar magnet (\ref{eq:Hamiltonian}) qualitatively. At sufficiently
high temperatures, the system is in a conventional (i.e., strongly disordered)
paramagnetic phase. Below the upper Griffiths temperature $T_u$, which is defined as the
transition temperature of a clean system having $J_z^\parallel \equiv J_u$, rare thick
slabs (``rare regions'') of strong ($J_z^\parallel = J_u$) layers show local magnetic
order, while the bulk is  nonmagnetic. Although individual such slabs are prevented from
developing true long-range order \cite{MerminWagner66}, they can undergo KT transitions
\cite{KosterlitzThouless73}. The unusual behavior, eqs.\ (\ref{eq:anom}) to
(\ref{eq:rho_s_perp}), is caused by the interplay between the randomness and the KT
physics of the rare regions. Ultimately, below the lower Griffiths temperature $T_l$ (the
transition temperature of a clean system having $J_z^\parallel \equiv J_l$), all layers
order magnetically, and the system exhibits the conventional (i.e., strongly ordered)
ferromagnetic phase.


We now use optimal fluctuation theory (i.e., Lifshitz-tail arguments \cite{Lifshitz64})
to derive the thermodynamics in the Griffiths region. The probability $w(L_{RR})$ for
finding a rare region of $L_{RR}$ consecutive strong layers reads
\begin{equation}
w(L_{RR}) \sim (1-c)^{L_{RR}} = e^{-\tilde c L_{RR} }~,
\label{Eq:w(L_RR)}
\end{equation}
with $\tilde c \equiv -\ln(1-c)$. Each individual such slab is equivalent to a 2D XY
model, and thus undergoes a KT transition at some thickness-dependent temperature
$T_{KT}(L_{RR})$. Finite-size scaling yields $T_u - T_{KT}(L_{RR}) \sim L_{RR}^{-1/\nu}$.
This result defines a cutoff length $L_c(T) \sim (T_u - T)^{-\nu}$.
 At any temperature $T<T_u$, all rare regions of thickness
$L_{RR}<L_c(T)$ are (locally) in the disordered phase, while those having $L_{RR}>L_c(T)$
are in the quasi long-range ordered KT phase.

Let us first consider a single rare region. According to KT theory
\cite{KosterlitzThouless73}, the spatial correlation function $C(\mathbf{x})$ in the KT
phase falls off as a power of the distance $|\mathbf{x}|$:
\begin{equation}
C(\mathbf{x}) \sim |\mathbf{x}|^{-\eta} \qquad (|\mathbf{x}| \to \infty)~.
\label{eq:KT-corr}
\end{equation}
The exponent $\eta$ is related to the renormalized (parallel) spin-wave stiffness
$\rho_{s,RR}$ of the slab via $\eta=T/(2\pi \rho_{s,RR})$. It takes the value $1/4$ at
the KT transition and is inversely proportional to $L_{RR}$ for very thick rare regions.
We thus model the thickness-dependence of $\eta$ via $\eta = \frac 1 4 L_c(T)/L_{RR}$,
which correctly describes both limits. The power-law correlations also lead to a
nonlinear magnetization-vs.-field curve within the KT phase,
\begin{equation}
m \sim H^{\eta/(4-\eta)}~, \label{eq:m-H-KT}
\end{equation}
which implies an infinite magnetic susceptibility.

We now combine the single-slab results, (\ref{eq:KT-corr}) and (\ref{eq:m-H-KT}), with
the size distribution (\ref{Eq:w(L_RR)}). We start with the response to an external
magnetic field $H$. Neglecting interactions between the rare regions for the moment, we
write the rare-region contribution to the magnetization as
\begin{equation}
m \sim  \int_{L_c(T)}^\infty dL_{RR} \, w(L_{RR}) \,
H^{\eta(L_{RR})/[4-\eta(L_{RR})]}~.
\label{eq:m-H-integral}
\end{equation}
For small fields, this integral can be evaluated using the saddle-point method. This
yields a saddle-point value $L_{sp}^2 = |\ln(H)| L_c(T) / (16 \tilde c)$, implying that
the response at small fields is dominated by thick rare regions. Inserting $L_{sp}$  into
(\ref{eq:m-H-integral}) immediately gives (\ref{eq:m-H}). This highly singular result
breaks down for $H>H_x \sim \exp[-16\tilde c L_c(T)]$, at which the saddle point reaches
$L_c(T)$. For $H \agt H_x$, the response is of the power-law type, $m \sim H^{1/15}$,
until the regular, linear-response part takes over at an even larger field $H_{\rm reg}
\sim \exp[-(15/14)\tilde c L_c(T)]$ \cite{MGNTV10}.

The parallel spin-wave stiffness $\rho_{s,\parallel}$ can be found analogously.
The free energy cost due to a twist of the BCs in either the $x$ or $y$
direction is simply a sum over all slabs in the KT phase. Each slab has the same
twisted BCs, thus, the total parallel stiffness is given by
\begin{equation}
\rho_{s,\parallel} \sim \int_{L_c(T)}^\infty dL_{RR} \, w(L_{RR})\,
\rho_{s,RR}(L_{RR})~.
\label{eq:rho_s_parallel-integral}
\end{equation}
This integral is dominated by the contribution near the lower limit, where $\rho_{s,RR}$
is approximately constant and equal to $2T/\pi$. To leading exponential accuracy, we thus
obtain $\rho_{s,\parallel} \sim \exp[-\tilde c L_c(T)]$, which leads to
(\ref{eq:rho_s_parallel}).

To discuss the perpendicular stiffness $\rho_{s,\perp}$, we apply twisted BCs in the $z$
direction. The resulting local twists occur mostly in the disordered bulk between the
rare regions; due to the randomness, they are not uniform but vary
from layer to layer. As the spatial positions of the rare regions are completely
random, the distribution of their nearest-neighbor distances $R$ is Poissonian, $P(R) =
R_{KT} \exp(-R/R_{KT})$, where $R_{KT} \sim \exp[\tilde c L_c(T)]$ is the typical
separation. The effective coupling between neighboring rare regions falls off
exponentially, $J_{\rm{eff}}^\perp (R) \sim \exp(-R/\xi_0)$, where $\xi_0$ is the bulk
correlation length. Combining this exponential form with $P(R)$
gives a power-law distribution for the effective couplings, i.e.,
\begin{equation}
\bar P (J_{\rm{eff}}^\perp) \sim (J_{\rm{eff}}^\perp)^{\frac 1 z -1}~.
\label{eq:J_eff-distrib}
\end{equation}
The Griffiths dynamical exponent $z\equiv R_{KT}/\xi_0$ takes the value $\infty$ at
$T_u$, and decreases with decreasing temperature. Writing the free energy due to the
twist of the BCs by an angle $\Theta$
  as $\Delta F  \sim \sum_z
J_{\rm{eff}}^\perp \, \Theta_z^2$, with $\sum_z \Theta_z = \Theta$,
and minimizing
w.r.t.\ the $\Theta_z$, we obtain \cite{MohanNarayananVojta10}
\begin{equation}
\rho_{s,\perp} \sim \langle 1/J_{\rm{eff}}^\perp \rangle^{-1}
\label{eq:rho_s_perp_av}
\end{equation}
where $\langle \cdots \rangle$ denotes the average over the distribution
(\ref{eq:J_eff-distrib}). This average diverges for $z>1$, implying
$\rho_{s,\perp}=0$
at temperatures just below $T_u$. Upon lowering $T$ further,
the exponent $z$ reaches the value 1 at a temperature
$T_s<T_u$. For $T<T_s$ (i.e., $z<1$), the average converges, yielding
a nonzero stiffness. Close to $z=1$, the average behaves as $\langle
1/J_{\rm{eff}}^\perp
\rangle \sim 1/(1-z)$ yielding (\ref{eq:rho_s_perp}).

Finally, we turn to the spontaneous magnetization $m$. The reason that $m>0$ for all
$T<T_u$ is the infinite susceptibility of those slabs that are in the KT phase.
They align to one another via an infinitesimal coupling. In contrast, in the quantum
Griffiths scenario, realized in the layered Heisenberg magnet
\cite{MohanNarayananVojta10}, the rare regions have a large but
finite susceptibility. Aligning them requires a nonzero coupling, so that long-range order
only appears at some critical temperature below $T_u$.
To estimate $m$, we combine the effective interaction $J_{\rm{eff}}^\perp$ with the KT
scaling within the rare regions. Consider an area of linear size $L$ (in the $x$ and $y$
directions) in one of the slabs. The typical magnetization (per site) of such
a region can be calculated by integrating (\ref{eq:KT-corr}),
yielding $m(L) \sim L^{-\eta/2}$. Now consider two such areas in neighboring rare
regions. Their interaction can be estimated as $J_{\rm{eff}}^\perp (L) =
J_{\rm{eff}}^\perp L^2 m^2(L) \sim L^{2-\eta} \exp(-R_{KT}/\xi_0)$. When this interaction
becomes of order $T$, the areas align, and long-range order sets in. This happens at a
length $L=L_x \sim [\exp(R_{KT}/\xi_0)]^{1/(2-\eta)}$, yielding
\begin{equation}
m \sim L_x^{-\eta/2} \sim \exp[-(R_{KT}/\xi_0)\eta/(4-2\eta)]~.
\label{eq:m-L_x}
\end{equation}
Because of the exponential size distribution (\ref{Eq:w(L_RR)}), the vast majority of
rare regions in the KT phase are very close  to the KT transition. Thus, to a good
approximation, we can set $\eta=1/4$. Inserting this, along with $R_{KT} \sim \exp[\tilde
c L_c(T)]$, into (\ref{eq:m-L_x}) yields the final result (\ref{eq:m-T}). This
calculation can be refined by taking into account the random distribution of rare-region
separations, which only modifies the nonuniversal constants in (\ref{eq:m-T})
\cite{MGNTV10}.


We now turn to the aspects of finite system size. The main
effect of a finite perpendicular size $L_\perp$, which is  experimentally important because
the number of layers in a real sample will often be small, is to limit the maximum
rare-region thickness $L_{RR}^{\rm{max}}$ in the sample. Estimating $L_{RR}^{\rm{max}}$ via
the condition that a sample of size $L_\perp$ contains, on average, exactly one such rare
region, i.e. $L_\perp w(L_{RR}^{\rm{max}}) \sim 1$, we obtain $L_{RR}^{\rm{max}} \sim
\ln(L_\perp) / \tilde c$.

We note that $L_{RR}^{\rm{max}}$ introduces an upper limit to the integral
(\ref{eq:m-H-integral}) for the $m(H)$ curve. When the saddle-point
value $L_{sp}$ is larger than $L_{RR}^{\rm{max}}$, which happens for fields $H < H_x$
with $\ln(H_x) \sim \ln^2(L_\perp)/[c L_c(T)]$, the integral is dominated by the
contribution near the upper limit. For very low fields, (\ref{eq:m-H}) gets thus replaced
by a power law with a size-dependent exponent: $m \sim H^{B \tilde c
L_c(T)/\ln(L_\perp)}$, with $B$ a constant.
The same mechanism also introduces an upper limit into the integral
(\ref{eq:rho_s_parallel-integral}) for the parallel stiffness. As
this integral is
dominated by the lower limit, the finite size only matters when
$L_c(T) > L_{RR}^{\rm{max}}$. Thus, the exponential tail
(\ref{eq:rho_s_parallel}) of $\rho_{s,\parallel}$ gets cut off near the upper
Griffiths temperature, for $T_u-T \lesssim [\ln(L_\perp)/\tilde
c]^{-1/\nu}$.
Using (\ref{eq:J_eff-distrib}), the minimum $J_{\rm eff}^\perp$ in a sample of size
$L_\perp$ behaves as $L_\perp^{-z}$. Inserting this into the elastic free energy
expression given above (\ref{eq:rho_s_perp_av}) yields the anomalous elasticity scaling
(\ref{eq:anom}).

As an example of the effects of a finite in-plane size $L_\parallel$, we discuss the
magnetic susceptibility. When $L_\parallel$ is finite, the susceptibility of a single
slab in the KT phase is no longer infinite. Its $L_\parallel$-dependence can be obtained
from integrating (\ref{eq:KT-corr}) to an upper cutoff $L_\parallel$, which yields
$\chi_{RR}(L_\parallel) \sim L_\parallel^{2-\eta}$. Summing this over all rare regions,
and evaluating the integral in the saddle-point approximation, gives a total
susceptibility (per unit volume) of $\chi \sim L_\parallel^2 \exp\{-[cL_c(T)
\ln(L_\parallel)]^{1/2}\}$.


In summary, we have shown that the randomly layered planar magnet features
anomalous elasticity and unusual thermodynamics in parts of the Griffiths phase.
%
Although we have considered the binary disorder distribution
(\ref{eq:binary_distrib}), the functional forms of the results (\ref{eq:anom}) to
(\ref{eq:rho_s_perp}) remain valid for any bounded distribution, provided it does not
vanish too rapidly at the upper bound. If the  distribution is unbounded, the tails of
magnetization and parallel stiffness would extend to $T=\infty$, implying that the system
is always in the magnetic phase \cite{MGNTV10}.

Our 
theory describes the regime where the system consists of
a few isolated rare regions in a disordered bulk; it becomes controlled for
$T\to T_u$. To describe the formation of bulk order close to $T_l$, the growths
and merging of rare regions need to be included. Moreover, the character of the
vortex unbinding transition changes for layers that are coupled to already
ordered slabs \cite{Fertig02,ZhangFertig05}.

The results (\ref{eq:anom}) to (\ref{eq:rho_s_perp}) have been formulated in terms
of the planar ferromagnet. Nonetheless, they apply to all transitions having $O(2)$ or
$U(1)$ order parameters, if expressed in terms of the appropriate variables. For layered
superfluids and superconductors \footnote{In superconducting multilayers, gauge fluctuations
introduce extra complications. The log.\ vortex interaction is cut-off or, at
least, weakened \cite{Babaev08} at a thickness-dependent length scale,  
limiting the rare region size.},
the magnetization should be exchanged for the condensate wave function or
the Cooper pair amplitude, respectively. In the same way, the spin-wave stiffness should
be exchanged for the superfluid density, and the external field could possibly be
realized via the proximity effect.

Let us relate our theory to the classification of phase transitions with
disorder based on the rare-region dimensionality $d_{RR}$
\cite{Vojta06,VojtaSchmalian05}:
It states that the critical behavior is conventional
if $d_{RR}$ is smaller than the lower critical dimension $d_c^-$ of the corresponding
clean transition; if the rare regions order independently (i.e., if $d_{RR} >
d_c^-$), the transition is smeared. The marginal case, $d_{RR}=d_c^-$,  usually leads to
an infinite-randomness critical point. Based on these arguments, one might expect
an infinite-randomness critical point in our system. However, the
\emph{quasi long-range} order that arises on rare regions in the  KT phase actually
leads to a \emph{hybrid} between a smeared and a sharp transition. On the one hand,
long-range order is present in the entire Griffiths phase (extending to $T=\infty$
for an unbounded disorder distribution), just as at the smeared transition of the
randomly layered Ising model \cite{Vojta03b}. On the other hand, the long-range order is
due to a collective effect (rather than individual freezing of rare regions), as in the
randomly layered Heisenberg magnet \cite{MohanNarayananVojta10}, which has a sharp
transition.

Not only are our results of conceptual importance for the theory of phase transitions,
but also they can be tested experimentally by producing layered nanostructures of
magnetic or superconducting materials. Magnetic multilayers having systematic variations
of $T_c$ from layer to layer have recently been produced \cite{MPHW09}, and our theory
should describe random versions of such structures (with XY spin symmetry). Moreover,
using ultra-cold atomic gases, one should be able to completely engineer the appropriate
many-particle Hamiltonian. We note that the Kosterlitz-Thouless transition in a single
slab of an $^{87}$Rb gas has already been observed \cite{HKCBD06}.


This work was supported in part by NSF awards DMR-0339147, DMR-0906566 and DMR-0906780,
and by the Research Corporation. We also acknowledge the hospitality of the Aspen Center of
Physics.

After completion of this work, we learned of a study of the same issues by means of a
numerical strong-disorder renormalization group \cite{PekkerRefaelDemler10}. Our phase
transition scenario agrees with that of Ref.\ \cite{PekkerRefaelDemler10}, and our
asymptotic analytical results complement their numerical data.

\bibliographystyle{apsrev}
\bibliography{../00Bibtex/rareregions}
\end{document}